\documentclass[preprint, nofootinbib]{revtex4}
\usepackage{amsmath, amsfonts, amssymb,graphicx, dcolumn}

\begin{document}

\title{Steady-State Two Atom Entanglement in a Pumped Cavity}
\author{Hideomi Nihira}
\email{nihira@optics.rochester.edu}
\author {C. R. Stroud, Jr.}
\affiliation{The Institute of Optics, University of Rochester,
Rochester, NY 14627, USA}
\date{\today}

\begin{abstract}

In this paper we explore the possibility of a steady-state
entanglement of two two-level atoms inside a pumped cavity by taking
into account cavity leakage and the spontaneous emission of photons
by the atoms.  We describe the system in the dressed state picture
in which the coherence is built into the dressed states while
transitions between the dressed states are incoherent. Our model
assumes the vacuum Rabi splitting of the dressed states to be much
larger than any of the decay parameters of the system which allows
atom-field coherence to build up before any decay process takes
over. We show that, under our model, a pumping field cannot entangle
two closed two-level atoms inside the cavity in the steady-state,
but a steady-state entanglement can be achieved with two open
two-level atoms.
\end{abstract}
\maketitle

\section{Introduction}

In this paper we investigate the steady-state entanglement of two
two-level atoms inside a high-Q cavity which is pumped by an
external field.  We assume a large vacuum Rabi splitting, and take
into account the presence of cavity leakage and spontaneous emission
by the atoms. We demonstrate that a pumping field cannot entangle
two two-level atoms inside a cavity in the steady-state, but it is
possible to have steady state entanglement between two open
two-level atoms in a cavity in the presence of a pumping field. This
is achieved by optically pumping half of the atomic population
outside of the decoherence-free subspace of the system, and the
process yields an entanglement equivalent to a Bell state content of
$\frac{1}{2}$.

Entanglement between two partitions of a system is often due to some
constraint placed on the dynamics of the system.  This constraint
can take the form of energy conservation, momentum conservation, or
structural constraint to name a few.  In this paper we direct our
attention to the entanglement that arises from the constraints on
the energy, or more specifically, the number of excitations in our
system.

In a simple system of two two-level atoms inside a cavity, initially
in their ground states, and with one photon in the cavity mode, we
know that the two atoms get entangled and disentangled periodically
in time.  This could lead one to think that if the two atoms were
placed in a sufficiently high-Q cavity with small losses introduced
and externally pumped such that the average photon number in the
cavity is less than or equal to one, then the two atoms would still
get entangled in some fashion.  In the simplest case in which one
pumps the cavity with a single photon source, and constrains the
pumping rate to be less than or equal to the decay rate of the
system, it would seem plausible that the two atoms inside the cavity
can be entangled.  In fact, making a weak field assumption and only
taking into account the first order correction to the reduced
density matrix of the atoms will yield a non-zero entanglement
between the atoms.  However, we show that if one were to take into
account the higher order corrections in the density matrix, then the
entanglement between the atoms vanish.  The result is due to the
fact that the concurrence measure is a nonlinear function of the
parameters of the atomic density matrix.  The weak field assumption
is only valid if one is calculating some property of the atoms which
is linearly dependent on the parameters.


Recent experimental advances in atomic traps and cavity QED
\cite{Wienman,Wineland,Haroche, Walther, Kimble} offer exciting new
possibilities in quantum computation and quantum networking using
the entanglement between atoms in the cavity \cite{Walther2}.  It is
within our technological limits to trap and manipulate atoms inside
a microcavity to study their entanglement behavior \cite{Haroche}.
Therefore, it is important to characterize the entanglement of the
atoms in the cavity, and how the entanglement is changed by the
dynamics of the system.

Theoretical work on atoms inside a cavity \cite{Guo, Li, Orozco,
Knight} suggests not only entanglement between atoms in a cavity,
but the possibility of a steady-state atomic entanglement in
presence of a weak pump. Here we will investigate under what
circumstances the atoms inside the cavity will get entangled with
each other, and whether a steady-state entanglement can be attained.

Our approach to characterizing the entanglement between atoms inside
a cavity is through the dressed state picture of the atom-field
states in the limit of large vacuum Rabi splitting.  This approach
enables us to use rate equations to characterize the transitions
between the dressed states \cite{Tan1, Stroud}.  All the coherent
effects are contained within the dressed states, and the transitions
between the dressed levels are incoherent effects which are
attributed to spontaneous decay of the atom, cavity leakage, or
pumping.  Also, for simplicity, we ignore longitudinal dipole-dipole
interaction between the atoms and motional effects in our
calculation.  The advantage to our approach, which expresses the
density matrix of the system as a mixture of the dressed states, is
that we can identify which dressed states contribute to the
entanglement between the atoms.  This will more directly suggest how
one may manipulate the system in order to maximize entanglement
between the atoms.  We will show that in the steady-state regime a
pumping field cannot entangle two closed two-level atoms in the
steady-state, but it is possible to entangle two open two-level
atoms in the steady-state with an entanglement content worth one
half a Bell state through optically pumping atomic population out of
the decoherence-free subspace of the system using a pumping field.

First, we will investigate the case in which two closed two-level
atoms are inside a cavity by taking into account the $n=0$, $n=1$,
and $n=2$ dressed states, and we show there is no entanglement
between the atoms.  Then we will show that a calculation involving
only the $n=0$ and $n=1$ dressed states yield a non-zero
entanglement between the atoms.  Although this approach is
reasonable for weak pumping fields when one is seeking the density
matrix for the purpose of determining the population distribution,
or other characteristics of the atoms which depend linearly on the
density matrix, we will show that taking into account only the $n=1$
dressed states is insufficient for the purposes of determining the
entanglement between the atoms.  We then proceed to show that a
steady-state entanglement can be obtained in the presence of a
pumping field for the case of two open two-level atoms even when we
consider all the dressed states of the system.

\section{Model System: Two Closed Two-level Atoms Inside a Cavity}

\subsection{Multiple Excitation in the Cavity}


In this section we will show that a pumping field cannot generate
entanglement between two two-level atoms inside a high-Q cavity in
the presence of cavity leakage and spontaneous emission by the
atoms.  The atoms are both initially in the ground state, and they
are placed in the cavity with the cavity mode being on resonance
with the atomic transition.  The external pumping field is also on
resonance with the atomic transition in our model. Furthermore, to
be consistent with the high-Q cavity assumption, we will stipulate
that the Rabi splitting is much larger than the spontaneous decay
rate and the cavity leakage rate.  The large Rabi splitting
justifies the rate equation approach we will take to analyze our
system.

\begin{figure}[h]
\centering
\includegraphics[width=5in]{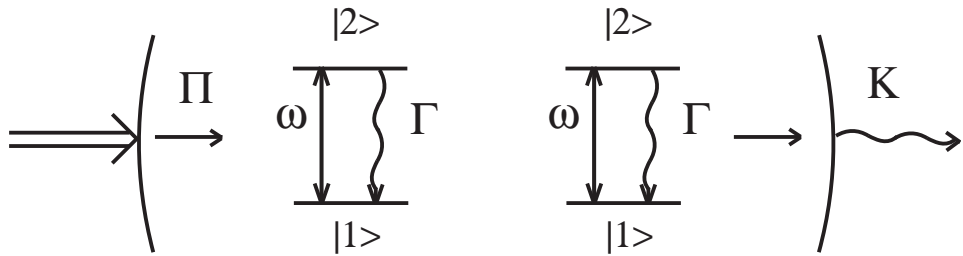}
\caption{\label{system} Two closed two-level atoms inside a cavity}
\end{figure}

The system we are considering consists of two atoms, each with
energy structure shown in Fig.(\ref{system}), in a cavity which is
externally pumped on resonance with the atomic transition. The
system can lose energy through cavity leakage or through spontaneous
emission of the atoms.

Since we wish to characterize this system using rate equations, we
will first obtain the dressed states of the closed system, and later
calculate the transition rates between these dressed states. We
assume that the Rabi splitting in this system is large such that the
coherence of the atoms and the field is contained in the dressed
states.  Therefore, pumping population from the ground state up to
these dressed states results in populating an incoherent mixture of
these dressed states. We will study the entanglement between the atoms in the steady-state regime
after we obtain the rate equations that govern
the population distribution of the system.

The Hamiltonian of the closed system in the interaction picture is

\begin{equation}
\hat{H}_{I}=g_{1}\hat{\sigma}_{12}^{(1)}\hat{a}^{\dagger}+g_{2}\hat{\sigma}_{12}^{(2)}\hat{a}^{\dagger}+H.C.
\end{equation}

\noindent where $\hat{\sigma}_{ij}^{(n)}$ is the atomic transition
operator for the $n^{th}$ atom and $\hat{a}^{\dagger}$ is the field
creation operator.  For the sake of simplicity we will assume the coupling constants, $g_{i}$, to be the same for each atom.  In the case in which there is only one excitation in the system there are three essential states,

\begin{equation}
\begin{array}{l}
|11;1\rangle \\
|12;0\rangle \\
|21;0 \rangle, \\
\end{array}
\end{equation}

\noindent and three dressed states,

\begin{equation}
|\chi_{o}\rangle=\frac{1}{\sqrt{2}}\big(|12;0\rangle-|21;0\rangle\big),
\end{equation}

\begin{equation}
|\chi_{+}\rangle=\frac{1}{\sqrt{2}}|11;1\rangle + \frac{1}{2} \big(|12;0\rangle+|21;0\rangle\big),
\end{equation}

\begin{equation}
|\chi_{-}\rangle=\frac{1}{\sqrt{2}}|11;1\rangle - \frac{1}{2} \big(|12;0\rangle+|21;0\rangle\big).
\end{equation}

\noindent Here $|ab;c\rangle=|a\rangle_{1}\otimes|b\rangle_{2}\otimes|c\rangle_{f}$ indicates the first atom is in state $|a\rangle$, the second atom in state $|b\rangle$, and the field in state $|c\rangle$.

In order to construct the rate equations governing the distribution
of the populations we will now need to calculate the spontaneous
emission rate, pumping rate, and the cavity leakage rate of the
system.  The spontaneous emission rate between states $|a\rangle$
and $|b\rangle$ is proportional to $|\langle
a|\hat{d}_{ab}|b\rangle|^{2}$ where $\hat{d}_{ab}$ is the dipole
matrix element between states $|a\rangle$ and $|b\rangle$.
Therefore, the rate of decay of the dressed states into the ground
state, $|g\rangle=|11;0\rangle$, is given by,

\begin{equation}
\begin{array}{l}
\gamma_{o\rightarrow g}=\Gamma|\langle \chi_{o}|\hat{d}^{(1)}+\hat{d}^{(2)}|g\rangle|^{2}=0,\\
\gamma_{+\rightarrow g}=\Gamma|\langle \chi_{+}|\hat{d}^{(1)}+\hat{d}^{(2)}|g\rangle|^{2}=\Gamma,\\
\gamma_{-\rightarrow g}=\Gamma|\langle \chi_{-}|\hat{d}^{(1)}+\hat{d}^{(2)}|g\rangle|^{2}=\Gamma,
\end{array}
\end{equation}

\noindent where $\Gamma$ is the Einstein A coefficient of the $2\rightarrow 1$ transition of the single atom in free space.

Likewise, the pumping rate of the cavity is proportional to $|\langle c|\hat{a}^{\dagger}|c-1 \rangle|^{2}$, where $|c\rangle$ is the field state in the Fock basis. The pumping rates for the dressed states from the ground state is given by,

\begin{equation}
\begin{array}{l}
\pi_{g\rightarrow o}=\Pi|\langle \chi_{o}|\hat{a}^{\dagger}|g\rangle|^{2}=0,\\
\pi_{g\rightarrow +}=\Pi|\langle \chi_{+}|\hat{a}^{\dagger}|g\rangle|^{2}=\frac{\Pi}{2},\\
\pi_{g\rightarrow -}=\Pi|\langle \chi_{-}|\hat{a}^{\dagger}|g\rangle|^{2}=\frac{\Pi}{2},
\end{array}
\end{equation}

\noindent where $\Pi$ is the single photon pumping rate inside the cavity (i.e. $\Pi=\Pi_{source}T^{2}$ where $T$ is the amplitude transmission coefficient of the input cavity mirror and $\Pi_{source}$ is the single photon emission rate of the pumping source).

The cavity leakage rate is obtained in a similar fashion,

\begin{equation}
\begin{array}{l}
\kappa_{o\rightarrow g}=K|\langle \chi_{o}|\hat{a}^{\dagger}|g\rangle|^{2}=0,\\
\kappa_{+\rightarrow g}=K|\langle \chi_{+}|\hat{a}^{\dagger}|g\rangle|^{2}=\frac{K}{2},\\
\kappa_{-\rightarrow g}=K|\langle \chi_{-}|\hat{a}^{\dagger}|g\rangle|^{2}=\frac{K}{2}.
\end{array}
\end{equation}

\noindent Here $K$ is the power transmission coefficient of the
cavity output mirror.

The $n\geq2$ excitation of the system will have a different set of
dressed states since there are four essential states in this case.
The four essential states for $n\geq2$ are,

\begin{equation}
\begin{array}{l}
|11;n\rangle \\
|12;n-1\rangle \\
|21;n-1 \rangle, \\
|22;n-2\rangle.\\
\end{array}
\end{equation}

\noindent The dressed states are then given by,

\begin{equation}
|\phi_{o}^{n}\rangle=\frac{1}{\sqrt{2}}\big(|12;n-1\rangle-|21;n-1\rangle\big),
\end{equation}

\begin{equation}
|\phi_{o'}^{n}\rangle=\frac{1}{\sqrt{2}}\big(|11;n\rangle-|22;n-2\rangle\big),
\end{equation}

\begin{equation}
|\phi_{+}^{n}\rangle=\frac{1}{2}\big(|11;n\rangle +
|12;n-1\rangle+|21;n-1\rangle+|22;n-2\rangle\big),
\end{equation}

\begin{equation}
|\phi_{-}^{n}\rangle=\frac{1}{2}\big(|11;n\rangle -
|12;n-1\rangle-|21;n-1\rangle+|22;n-2\rangle\big).
\end{equation}

After we calculate the pumping rate, cavity leakage rate, and
spontaneous emission rate by the atoms between the dressed states,
as we have done before for the $n=1$ dressed states and the ground
state, we can construct the rate equation governing the population
distribution amongst the $n=0$, $n=1$, and $n=2$ dressed states.
The rate equation is given by,

\begin{equation}
\begin{array}{l}
\dot{P}_{g}=\Gamma P_{s1}+\frac{K}{2}P_{s1}-\Pi P_{g},\\
\dot{P}_{s1}=-\Gamma P_{s1}-\frac{K}{2}P_{s1}+\Pi P_{g}+\frac{3}{2}\Gamma P_{s2}+\Gamma P_{o',2}+K(P_{s2}+P_{o',2})+\frac{3}{2}\Pi P_{s1},\\
\dot{P}_{s2}=-\frac{3}{2}\Gamma P_{s2}-KP_{s2}+\Pi P_{s1},\\
\dot{P}_{o',2}=-\Gamma P_{o',2}-KP_{o'}+\frac{1}{2}\Pi P_{s1},\\
\dot{P}_{o}=0,\\
\dot{P}_{o,2}=0,\\
P_{s1}=P_{+}+P_{-},\\
P_{s2}=P_{+,2}+P_{-,2},
\end{array}
\end{equation}

\noindent where $P_{g}$ is the population of the ground state of the
system, $P_{\pm}$ is the population in the $|\chi_{\pm}\rangle$
dressed states, $P_{0}$ is the population in the $|\chi_{o}\rangle$
dressed state, $P_{\pm,n}$ is the population in
$|\phi_{\pm}^{n}\rangle$, $P_{o,n}$ is the population in the
$|\phi_{o}^{n}\rangle$, and $P_{o',n}$ is the population in
$|\phi_{o'}^{n}\rangle$.  Here we assume that there is no population
initially in the $|\chi_{o}\rangle$ and $|\phi_{o}^{n}\rangle$
dressed states, and because these states don't couple to any other
states, they will not accumulate any population at later times.

The steady-state solution to these rate equations is,

\begin{equation}
\begin{array}{l}
P_{g}=\mathcal{N}(3\Gamma+2K)(2\Gamma+K)(\Gamma+K),\\
P_{s1}=\mathcal{N}2\Pi(\Gamma+K)(3\Gamma+2K),\\
P_{s2}=\mathcal{N}4\Pi^{2}(\Gamma+K),\\
P_{o',2}=\mathcal{N}\Pi^{2}(3\Gamma+2K),\\
\mathcal{N}=7\Gamma\Pi^{2}+6\Pi^{2}K+6\Pi\Gamma^{2}+10\Pi K\Gamma+4\Pi K^{2}+6\Gamma^{3}+13K\Gamma^{2}+9K^{2}\Gamma+2K^{3}.\\
\end{array}
\end{equation}

Are the two atoms in the cavity entangled?  In order to answer this
question we will have to calculate the entanglement content of this
bipartite two-level mixed state of the system.  Here we will employ
the concurrence measure put forward by Wootters \cite{wooters}.  The
concurrence measure of entanglement is defined as,

\begin{equation}
C=max(\sqrt{\lambda_{1}}-\sqrt{\lambda_{2}}-\sqrt{\lambda_{3}}-\sqrt{\lambda_{4}},0)
\end{equation}

\noindent where $\lambda_{i}$ are the eigenvalues, in descending
order of value, of the matrix $\rho\tilde{\rho}$
($\tilde{\rho}=(\sigma_{y}\otimes\sigma_{y})\rho^{*}(\sigma_{y}\otimes\sigma_{y})$).

To determine the entanglement between the atoms in the cavity we
will trace out the field component of the density matrix to obtain
the reduced density matrix of the system.  The reduced density
matrix of the two atoms is given by,

\begin{equation}
\hat{\rho}_{atoms}=P_{g}\hat{\rho}_{g}+P_{s1}\hat{\rho}_{s1}+P_{s2}\hat{\rho}_{s2}+P_{o',2}\hat{\rho}_{o',2}
\end{equation}

\noindent where

\begin{equation}
\hat{\rho}_{g}= \begin{pmatrix}
1 & 0 & 0 & 0 \\
0 & 0 & 0 & 0 \\
0 & 0 & 0 & 0 \\
0 & 0 & 0 & 0 \\
\end{pmatrix} ,
\hat{\rho}_{s1}= \begin{pmatrix}
\frac{1}{2} & 0 & 0 & 0 \\
0 & \frac{1}{4} & \frac{1}{4} & 0 \\
0 & \frac{1}{4} & \frac{1}{4} & 0 \\
0 & 0 & 0 & 0 \\
\end{pmatrix},
\hat{\rho}_{s2}= \begin{pmatrix}
\frac{1}{4} & 0 & 0 & 0 \\
0 & \frac{1}{4} & \frac{1}{4} & 0 \\
0 & \frac{1}{4} & \frac{1}{4} & 0 \\
0 & 0 & 0 & \frac{1}{4} \\
\end{pmatrix},
\hat{\rho}_{o',2}= \begin{pmatrix}
\frac{1}{2} & 0 & 0 & 0 \\
0 & 0 & 0 & 0 \\
0 & 0 & 0 & 0 \\
0 & 0 & 0 & \frac{1}{2} \\
\end{pmatrix}.
\end{equation}

The important point to note in the dressed states for $n\geq2$ is
that, aside from the $|\phi_{o}^n\rangle$ which does not couple to
any of the other dressed states through spontaneous emission,
pumping, or leakage, the other three dressed $|\phi_{o'}^n\rangle$,
$|\phi_{+}^n\rangle$, and, $|\phi_{-}^n\rangle$ have no entanglement
(i.e $C=0$) between the atoms.  This is in contrast with the $n=1$
dressed states, $|\chi_{\pm}\rangle$, which does contain
entanglement between the two atoms ($C=\frac{1}{2}$).  This suggests
that any strong pumping which puts appreciable population in the
$n\geq2$ will disentangle the atoms.  On the other hand, it may seem
to suggest that in the limit of weak pumping, in which the $n\geq2$
dressed states are not appreciably populated, the atoms may turn out
entangled.  We will show that the atoms do not get entangled no
matter how weak the pumping field.

 The concurrence of $\hat{\rho}_{atoms}$ is,

\begin{equation}
C_{atoms}=max\big(\frac{1}{2}(P_{s1}+P_{s2})-\frac{1}{2}\sqrt{(P_{s2}+2P_{o',2})(2P_{o',2}+P_{s2}+4P_{g}+2P_{s1})},0\big).
\end{equation}

\begin{figure}[h]
\centering
\includegraphics[width=5in]{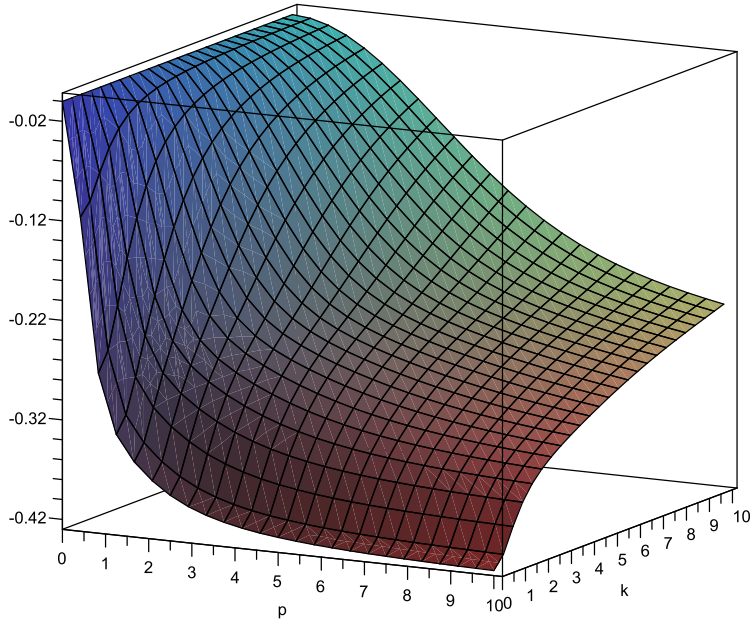}
\caption{\label{Concurrence} Plot of $\mathcal{C}$ against $\Pi$ and
$K$ (in units of $\Gamma$)}
\end{figure}

\noindent  It is not easy to determine by observation if there
exists some combination of parameters which would provide a non-zero
concurrence. The plot of the expressions inside the $max$ function
of the concurrence (which we will call $\mathcal{C}$) against $\Pi$
and $K$ (in units of $\Gamma$) Fig. (\ref{Concurrence}) suggests
that there is indeed no combination of parameters which will provide
a non-zero entanglement content.

Can there be a miniscule amount of entanglement that we don't see in
the plot?  In order to answer this let us take a closer look at the
expression for concurrence.  First, we will determine if there are
any extremums in the expression for concurrence by looking at the
partial derivatives of $C$ with respect to the $P_i$ parameters and
setting the expression to zero.  After differentiating C we obtain,

\begin{equation}
\begin{array}{l}
\frac{\partial C}{\partial P_g}=-\frac{\mathcal{A}}{4}(8P_{o',2}+4P_{s2})=0,\\
\frac{\partial C}{\partial P_{s1}}=\frac{1}{2}-\frac{\mathcal{A}}{4}(4P_{o',2}+2P_{s2})=0,\\
\frac{\partial C}{\partial P_{s2}}=\frac{1}{2}-\frac{\mathcal{A}}{4}(4P_g+2P_{s1}+2P_{s2}+4P_{o',2})=0,\\
\frac{\partial C}{\partial P_{o',2}}=-\frac{\mathcal{A}}{4}(4P_{s2}+8P_g+4P_{s1}+8P_{o',2})=0,\\
\mathcal{A}=\frac{1}{\sqrt{4P_{s2}P_{g}+2P_{s1}P_{s2}+P_{s2}^2+4P_{s2}P_{o',2}+8P_{o',2}P_{g}+4P_{o',2}P_{s1}+4P_{o',2}^2}}.
\end{array}
\end{equation}

\noindent It is easy to see that $\frac{\partial C}{\partial P_g}=0$
and $\frac{\partial C}{\partial P_{o',2}}=0$ cannot be satisfied by
positive real values of the $P_i$.  The $P_i$ have to be positive
real values since they are the probability weighting of the density
matrix, and the steady-state solution suggests a non-zero value for
all the $P_i$.  The $\frac{\partial C}{\partial P_{s1}}=0$ and
$\frac{\partial C}{\partial P_{s2}}=0$ takes some algebra, but one
can also show that the equations cannot be satisfied within the
restrictions of the parameters.  Therefore, there are no extrema or
saddle points in the concurrence, which is consistent with Fig.
(\ref{Concurrence}).  This means that the extremum value lies on the
boundary of the domain of the $P_i$. The four boundary points are
$(P_g=1,P_{s1}=0,P_{s2}=0,P_{o',2}=0)$,
$(P_g=0,P_{s1}=1,P_{s2}=0,P_{o',2}=0)$,
$(P_g=0,P_{s1}=0,P_{s2}=1,P_{o',2}=0)$, and
$(P_g=0,P_{s1}=0,P_{s2}=0,P_{o',2}=1)$, of which only the
$(P_g=0,P_{s1}=1,P_{s2}=0,P_{o',2}=0)$ point offers a nonzero
concurrence of $C=\frac{1}{2}$.  Hence, the only possible non-zero
value of the concurrence is along the $P_{s1}$ axis from
$(P_g=0,P_{s1}=1,P_{s2}=0,P_{o',2}=0)$.  Our task is to determine
the maximum value of $P_{s1}$ constrained by the steady-state
solutions to the rate equations.

\begin{figure}[h]
\centering
\includegraphics[width=5in]{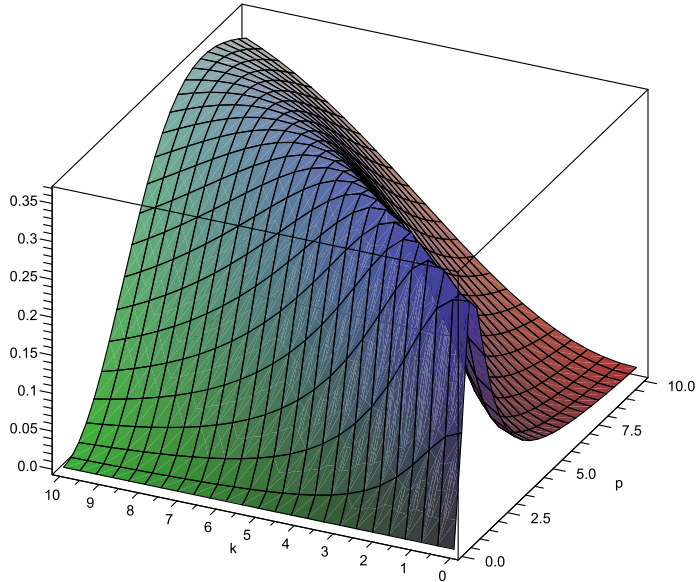}
\caption{\label{Pop} Plot of $P_{s1}$ against $\Pi$ and $K$ (in
units of $\Gamma$)}
\end{figure}

Using numerical techniques one can determine the maximum value of
$P_{s1}$ to be $P_{s1}\approx0.366$ which implies $P_g\approx0.317$,
$P_{s2}\approx0.211$, and $P_{o',2}\approx0.106$ (see Fig.
(\ref{Pop})).  These population yields a concurrence of
$C=max(\sqrt(0.083)-2\sqrt(0.0640),0)=0$.  This means that a pumping
field that takes any population beyond the $n=1$ dressed state
leaves the atoms completely unentangled.

There are a couple points here worth noting.  First, we only
extended the rate equations to include the $n=2$ dressed states, and
yet the concurrence suddenly dropped to zero.  In reality, the $n>2$
dressed states do get populated as well, so as far as the
entanglement of the atoms are concerned we are actually
overestimating the concurrence.  This is because in reality only
some fraction of that population resides in the $n=2$ states and the
rest is somewhere in the $n>2$ dressed states.  Therefore, this
implies even less population in the $|\chi_{\pm}\rangle$ dressed
states, and, as a consequence, less entanglement between the atoms.
Second, we would expect that in the limit of a weak pump the $n=2$
dressed states would not get appreciably populated, so the
entanglement between the atoms ought to be non-zero.  However, our
analysis indicates that any population at all in the $n=2$ dressed
states disentangles the atoms.  Lets take a numerical example to
illustrate this peculiar point.  For $\Pi=.447$ and $K=1$ (in units
of $\Gamma)$ we get $P_g=0.97337$, $P_{s1}=0.02595$,
$P_{s2}=0.00042$, and $P_{o',2}=0.00026$.  One would be tempted to
neglect the $P_{s2}$ and $P_{o'2}$ terms in the density matrix since
they are two orders of magnitude smaller than $P_{s1}$.  If we only
keep the leading two terms then we get a concurrence of $C=.01$.
However, if we were to keep all the terms in the density matrix we
would get a concurrence of $C=0$.  Although $P_{s2}$ and $P_{o'2}$
were two orders of magnitude smaller, by keeping these terms in the
density matrix we see there is no entanglement between the atoms. In
principle, we can always extend the accuracy of our model by
increasing the Rabi splitting of the dressed states.  Therefore, the
miniscule population in $n=2$ disentangling the atoms is indeed the
case for the model we are considering.  Because concurrence is not
necessarily linear in the parameters of the density matrix, one
cannot ignore terms even when they are several orders of magnitude
smaller.  In order to get a non-zero entanglement between the atoms
we have to put more population in the $|\chi_{\pm}\rangle$ dressed
states, and minimize the population transfer to the $n\geq2$ dressed
states.  The only realistic option we have is to have cavity leakage
rate that is nonlinear in the photon number \cite{Gaeta, Stank}
since $\Gamma$ is a fixed parameter of the atoms and $\Pi$ is a
pumping parameter that does not differentiate between the cavity
field excitation number.  If the mirror of the cavity demonstrates a
strong enough nonlinearity between the one excitation and two
excitation of the cavity field then one can obtain a non-zero
concurrence between the two atoms.


Finally, we would like to offer a more intuitive justification for
our use of the rate equations.  Clearly, the way we construct the
rate equations suggests that the populations of the dressed states
depends only on the populations of the other dressed states and not
the coherence between the dressed states.  The downward transitions,
spontaneous emission of the atoms and cavity decay, are incoherent
processes, so they do not impart any coherence onto the system.  The
upward transition, the external pumping, also does not impart any
coherence in the steady-state limit.  When a photon impinges on an
atom initially in the ground state inside a cavity, the system is
put in a coherent superposition of the dressed states whose
amplitudes oscillate with the associated Rabi frequency.  However,
when a decay process is introduced, the atom-cavity system can lose
the photon and go back into the ground state.  At this point they
wait for the next pump photon to come and restart the Rabi
oscillation.  In effect, the quantum jump that is introduced
randomizes the initial phase of the dressed states Rabi oscillation.
We do not know when the photon was lost nor do we know when the new
pump photon has arrived.  Therefore, in the longtime limit, after
many decay and pumping processes have taken place, we do not expect
to have any coherence between the dressed states since we will have
no information concerning the phase of the Rabi oscillation.  For
this reason we can use rate equations to describe our system.

\subsection{Single Excitation in the Cavity}

In the previous section we demonstrated that a pumping field cannot
entangle the atoms in the steady-state inside the cavity under our
model. In this section we will briefly exam what one might obtain
for the entanglement between the atoms if one employs the weak field
assumption and decides to only consider the $n=1$ dressed states.
The model seems plausible in the limit of a weak pumping field, but
we will show how the weak pumping assumption does not translate to
only considering the $n=1$ dressed states.

If one were to consider only the $n=1$ dressed states then the rate
equation governing the population distribution given by,

\begin{equation}
\begin{array}{l}
\dot{P}_{g}=\Gamma(P_{+}+P_{-})+\frac{K}{2}(P_{+}+P_{-})-\Pi P_{g},\\
\dot{P}_{o}=0,\\
\dot{P}_{+}=-\Gamma P_{+}-\frac{K}{2}P_{+}+\frac{\Pi}{2}P_{g},\\
\dot{P}_{-}=-\Gamma P_{-}-\frac{K}{2}P_{-}+\frac{\Pi}{2}P_{g}.\\
\end{array}
\end{equation}

\begin{figure}[h]
\centering
\includegraphics[width=4in]{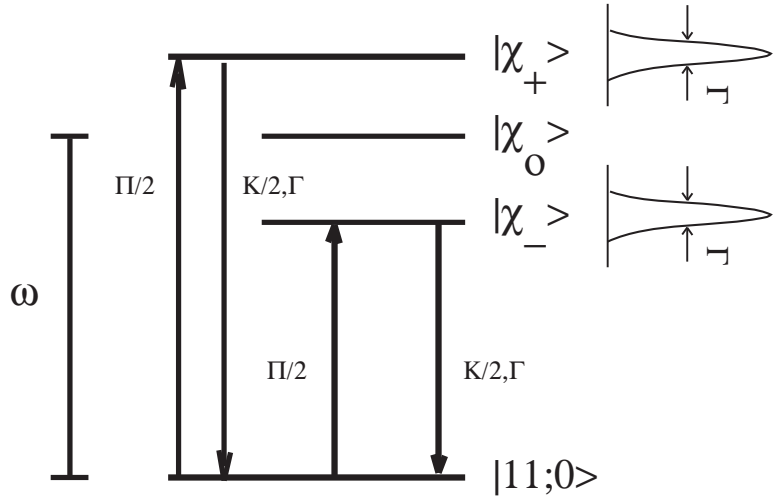}
\caption{\label{dressed} Schematic of the dressed state}
\end{figure}

\noindent Provided there is no population in $P_{o}$ at $t=0$, the
steady-state solution of these equations is,

\begin{equation}
\begin{array}{l}
P_{+}=P_{-}=N \frac{\Pi}{2},\\
P_{g}=N(\frac{K}{2}+\Gamma),\\
N = \frac{1}{\Gamma+\frac{K}{2}+\Pi}.
\end{array}
\end{equation}

The total density matrix (including the two atoms and the field) is
given by,

\begin{equation}
\hat{\rho}=P_{g}|g\rangle\langle g|+P_{+}|\chi_{+}\rangle\langle\chi_{+}|+P_{-}|\chi_{-}\rangle\langle\chi_{-}|+P_{o}|\chi_{o}\rangle\langle\chi_{o}|.
\end{equation}

\noindent Since we are only interested in the entanglement between the two atoms, we will trace out the field part of the density matrix to obtain the reduced density matrix for the two atoms.  This is given by,

\begin{equation}
Tr_{field}(\hat{\rho})=\big(P_{g}+\frac{1}{2}(P_{+}+P_{-})\big) |11\rangle\langle 11|+\frac{1}{2}(P_{+}+P_{-})|\psi^{+}\rangle\langle\psi^{+}|+P_{o}|\psi^{-}\rangle\langle\psi^{-}|
\end{equation}

\noindent where $|\psi^{+}\rangle=\frac{1}{\sqrt{2}}\big(|12\rangle+|21\rangle\big)$ and $|\psi^{-}\rangle=\frac{1}{\sqrt{2}}\big(|12\rangle-|21\rangle\big)$ are the two Bell states.  Substituting our previous steady-state solution to the above equation yields,

\begin{equation}
\hat{\rho}_{atoms}=N \big[ (\frac{\Pi+K}{2}+\Gamma)|11\rangle\langle 11|+\frac{\Pi}{2}|\psi^{+}\rangle\langle\psi^{+}|\big].
\end{equation}

The concurrence for this reduced density matrix is simply
$C=N\frac{\Pi}{2}$.  This model suggests that whenever the pump is
on there is always some entanglement between the atoms inside the
cavity.  This result is contrary to the one we have obtained
earlier.  If we only consider the $n=1$ dressed states, then there
is always some entanglement between the two atoms in the
steady-state when the pump is on. However, we have shown earlier
that, no matter how weak the pumping field, the two atoms do not get
entangled in the steady-state. The discrepancy between the two
results is due to the fact that concurrence is not a linear function
of the density matrix.  If concurrence were linearly related to the
density matrix, then the weak field assumption would be valid, and
the two atoms would indeed be entangled.  But because concurrence is
not a linear function of the density matrix (or, the wave function
of the system), one cannot dismiss the $n\geq2$ terms in the density
matrix even when the population in those states are significantly
smaller than the $n=1$ and $n=0$ dressed states. Therefore, the weak
field assumption does not translate to just keeping the lowest order
terms in the density matrix if one is interested in calculating the
entanglement content of a system.

In our analysis we have assumed that the atoms were on resonance
with the cavity field mode and the coupling constants of the atoms
were the same.  If the atoms had a different coupling constant
$g_{i}$, or detuned from the cavity field by $\Delta_{i}$, this
would result in the reduction of the entanglement between the atoms.
The entanglement between the atoms here is due to the
$|\psi^{+}\rangle$ in the $|\chi_{\pm}\rangle$ dressed states.
Different coupling constants or detuning would turn the
$|\psi^{+}\rangle$ into
$\frac{1}{\sqrt{|\alpha|^2+|\beta|^2}}\big(\alpha|12\rangle+\beta|21\rangle\big)$
in the dressed states, and in general, $\alpha\neq\beta$.  Because
maximum entanglement is attained when
$\alpha=\beta=\frac{1}{\sqrt{2}}$, the effect of differing $g_{i}$'s
or $\Delta_{i}$'s would reduce the entanglement between the atoms.
Therefore, the case we are considering is the case for maximum
possible entanglement between the atoms.

\section{Model System: Two Open Two-level Atoms Inside a Cavity}

\begin{figure}[h]
\centering
\includegraphics[width=5in]{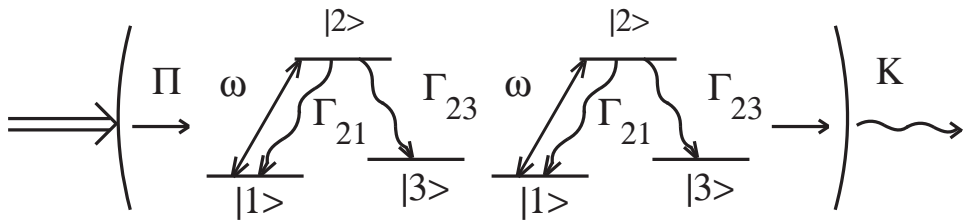}
\caption{\label{threelev} Two open two-level atoms inside a cavity}
\end{figure}

So, is there a way to generate steady-state entanglement with
pumping/leakage through the cavity and spontaneous emission?  Indeed
there is a way to generate entanglement under these conditions using
two open two-level atoms.  The system we consider,
Fig.(\ref{threelev}), is two open two-level atoms inside a cavity
with spontaneous emission down to both ground states and the cavity
being in resonance with only one transition of the open two-level
atom.

Since the pumping field does not interact with $|3\rangle$, the
Hamiltonian of the closed system is the same as the two closed two-level
system model, and therefore the dressed states are also the same.
The only difference is that when we construct our rate equations we
have the possibility of the atom to fall into the $|3\rangle$ state
through spontaneous emission.

Now, if we were to start out in the $|11;0\rangle$ state as we did
before then we will get no entanglement between the atoms since the
$|33;n\rangle$ state is clearly separable, and we saw in the
previous section that the population distribution in the dressed
states yields a concurrence of zero.  Here we will take advantage of
the maximally entangled $|\chi_{o}\rangle$ and $|\phi_{o}^n\rangle$
states by setting the initial state of the two atoms to be a linear
combination of the symmetric and antisymmetric states.  Assuming we
start out in the $|11;0\rangle$ state, we will apply a $\pi$-pulse
to one of the atoms before we turn the pumping field on.  This will
take the $|11;0\rangle$ state to the $|12;0\rangle$ state.  In terms
of the dressed states this is expressed as,

\begin{equation}
|12;0\rangle=\frac{1}{2}(|\chi_+\rangle -|\chi_-\rangle)+\frac{1}{\sqrt{2}}|\chi_o\rangle.
\end{equation}

\noindent If the atoms are left alone in free space in the $|12;0\rangle$ state then the state will decay down to,

\begin{equation}
\hat{\rho}_{free space}=\frac{\Gamma_{21}}{\Gamma_{21}+\Gamma_{23}}|11;0\rangle\langle 11;0|+\frac{\Gamma_{23}}{\Gamma_{21}+\Gamma_{23}}|13;0\rangle\langle 13;0|
\end{equation}

\noindent which is an unentangled mixed state of the system.  In
fact, in free space the dressed state picture offers no advantage
since the Rabi splitting is so small that the line widths of the
dressed states overlap, and consequently the dressed states don't
decay independently of each other.  However, when the two atoms are
placed in a cavity in which the Rabi splitting is larger than the
line widths, the dressed states do not overlap.  This gives rise to
the dressed states $|\chi_o\rangle$ and $|\phi_{o}^n\rangle$ that do
not couple to any of the other dressed states but themselves.  Note
that $|\chi_o\rangle$ and $|\phi_{o}^n\rangle$ both have the same
concurrence, $C=1$, and that the reduced density matrix of both
these states are the same, namely,

\begin{equation}
Tr_{field}(|\chi_o\rangle\langle \chi_o|)=Tr_{field}(|\phi_{o}^n\rangle\langle \phi_{o}^n|)=\frac{1}{2}(|12\rangle-|21\rangle)(\langle 12|-\langle 21)\equiv |\psi^-\rangle\langle \psi^-|.
\end{equation}

The idea of starting out in the $|12;0\rangle$ state is to maintain
the population in $|\chi_o\rangle$($|\phi_{o}^n\rangle$) component
of the density matrix, and pumping the rest of the population out of
the $|\chi_{\pm}\rangle$ and $|\phi_{\pm}^n\rangle$ states.  Why do
we want to pump the population out of the $|\chi_{\pm}\rangle$
states when it has a $C=\frac{1}{2}$?  The entanglement content of
$|\chi_{\pm}\rangle$ comes from the $|\psi^+\rangle$ Bell state
component.  Unfortunately, the incoherent mixture of different Bell
state components degrade the entanglement, and in the worst case
make the concurrence zero.  The extreme case is,

\begin{equation}
\hat{\rho}_{atoms}=\frac{1}{2}(|\psi^+\rangle\langle \psi^+|+|\psi^-\rangle\langle \psi^-|)=\frac{1}{2}(|12\rangle\langle 12| + |21\rangle\langle 21|)
\end{equation}

\noindent which is clearly a separable state.  Therefore, to
preserve the entanglement in the
$|\chi_o\rangle$($|\phi_{o}^n\rangle$) state we want to minimize the
population in the $|\chi_{\pm}\rangle$ and $|\phi_{\pm}^n\rangle$
states which contain a $|\psi^+\rangle$ component.  The system we
put forward does indeed pump the population out of the
$|\chi_{\pm}\rangle$ and $|\phi_{\pm}^n\rangle$ states and puts the
population in the $|33;n\rangle$ state.  Once the two atoms prepared
in the $|12;0\rangle$ state is put in the cavity the rate equations
governing the population in the $|\chi_{\pm}\rangle$ and
$|\phi_{\pm}^n\rangle$ are the same as the ones we have derived in
the previous section except they will contain a decay term that
would put the population in the $|33;n\rangle$ state.  Since the
pumping field is not in resonance with the $|3\rangle
\leftrightarrow |2\rangle$ transition the population in the
$|33;n\rangle$ state will not get pumped out.  In the steady-state,
the reduced density matrix of the atoms would then be,

\begin{equation}
\hat{\rho}_{atoms}=\frac{1}{2}|33\rangle\langle 33| + \frac{1}{2}  |\psi^-\rangle\langle \psi^-|.
\end{equation}

This is a three level state, so we cannot use Wootters' concurrence
to calculate the entanglement content, but a simple argument can
convince us that this state is entangled with a Bell state content
of one half.  Let's consider a very crude distillation protocol in
which we project $\hat{\rho}_{atoms}$ onto the $|33\rangle$ state.
Clearly, half the time we will detect the state to be in
$|33\rangle$, but the other half of the time we will get a null
result in our detector suggesting that the atoms are in
$|\psi^-\rangle$ state.  This means that we can obtain a
$|\psi^-\rangle$ Bell state half of the time, and therefore the
state contains at least a Bell state content of one half.  In this
case, we can go a step further and say because the mixture has half
its population in the Bell state, one cannot distill more than a
half a Bell state for this density matrix.  Therefore, our
distillation protocol is indeed optimum, and there is a Bell state
content of one half in this state. Another point to note here is
that the final state is independent of the pumping, leakage, and
spontaneous decay parameters in the rate equations. These parameters
only determine the time scale in which the atoms reach the
steady-state, but not the final state.

Now we can ask ourselves what would happen if we started out in the
$|12;0\rangle$ state for the two closed two-level atoms we have
considered in the previous section.  The steady-state reduced
density matrix of the two closed two-level atoms in the cavity would
be,

\begin{equation}
\hat{\rho}_{atoms}=\frac{1}{2} |\psi^-\rangle\langle
\psi^-|+\frac{1}{2}\Big(P_{g}\hat{\rho}_{g}+P_{s1}\hat{\rho}_{s1}+P_{s2}\hat{\rho}_{s2}+P_{o',2}\hat{\rho}_{o',2}\Big).
\end{equation}

\noindent The concurrence of this density matrix is given by,

\begin{equation}
C_{atoms}=max\big(\frac{1}{2}-\frac{1}{4}(P_{s1}+P_{s2})-\frac{1}{4}\sqrt{(P_{s2}+2P_{o',2})(2P_{o',2}+P_{s2}+4P_{g}+2P_{s1})},0\big).
\end{equation}

This suggests that the concurrence of the closed two-level atoms can
indeed be non-zero, and the concurrence is maximum ($C=\frac{1}{2}$)
when all the population that is not in the $|\phi_{o}^n\rangle$ or
$|\chi_{o}\rangle$ is in the $|11;0\rangle\langle 11;0|$ state.
Therefore, the concurrence in this case is maximum when the cavity
is not pumped at all.  As we have noted before, the Bell state
component that is generated through the pumping process is the
$|\psi^+\rangle$ state, and the incoherent mixture of
$|\psi^+\rangle$ and $|\psi^-\rangle$ degrades the entanglement
between the atoms.  Therefore, by pumping into the $|\psi^+\rangle$
state the concurrence is reduced.  This is consistent with the above
expression for the concurrence between the two atoms.  It follows that
the entanglement generated between the atoms in both the open and
closed two-level system is due to the asymmetric initial state and
the presence of the cavity.  The pumping itself does not directly
contribute to the entanglement.  What the pumping does in the open
two-level system is that the pumping field puts half the population
in the $|33;n\rangle$ state of the system to maximize the
entanglement between the atoms.  The advantage of the open two-level
atoms is that if one were to perform operations on the two-atoms in
the $(|1\rangle,|2\rangle)$ subspace, then the $|3\rangle$ state
would not interfere with the operation while in the case of the
close two-level atoms one could not easily determine if the
operation was performed on the entangled or the unentangled
component of the density matrix.

\section{Conclusion}

We have shown that in our model for two atoms inside a cavity a
pumping field cannot entangle two closed two-level atoms in the
steady-state, but it is possible to have entanglement in the
presence a pumping field for two open two-level atoms with a Bell
state content of one half through optically pumping atomic
population out of the decoherence-free subspace of the system. We
assume a situation in which the vacuum Rabi splitting is larger than
any of the decay parameters in the model, and we do not consider
dipole-diploe interactions or motional effects in our calculation.
The large Rabi splitting suggests the atoms and field can build up
coherence before any decay process takes over. Our assumption allows
us to treat the incoherent process of population transfer between
the dressed state of the system with rate equations and account for
the coherence between the atoms through the dressed states. This
approach explicitly points out the states that contribute to the
entanglement of the atoms (the $n=1$ dressed states and the
anti-symmetric dressed states). Qualitatively, one can see that if
there are $m$ atoms in the cavity then the dressed states associated
with $n<m$ excitations are the states that contain some entanglement
between the atoms.  Because of the various types of non-equivalent
entanglement for $m>2$ partition systems, there are no standard
measures for entanglement and the analysis is beyond the scope of
this paper.  However, if one is only looking for a particular type
of entanglement present in the system, then we believe that the
dressed state picture we present offers a more direct way to see
where the target states lie in the dressed state ladder.

\end{document}